# A Reputation Economy:

# Results from an Empirical Survey on Academic Data Sharing

**Benedikt Fecher** [a, b,*], **Sascha Friesike** [b], **Marcel Hebing** [c],
**Stephanie Linek** [d], **Armin Sauermann** [b]

[a] Department "Research Infrastructure", German Institute for Economic Research (DIW Berlin), Berlin, Germany

[b] Department "Internet-Enabled Innovation", Alexander von Humboldt Institute for Internet and Society, Berlin, Germany

[c] German Socio-Economic Panel Study, German Institute for Economic Research (DIW Berlin), Berlin, Germany

[d] Department "Science 2.0", German National Library of Economics (ZBW), Kiel, Germany

* Corresponding author

Berlin and Kiel, February 2015

**Acknowledgement**

We thank Gert G. Wagner and Mathis Fräßdorf for their helpful comments and suggestions. We thank Klaus Tochtermann and the Leibniz Science 2.0 Research Association for support. We gratefully acknowledge the financial support from the German Research Foundation (DFG) under the project "European Data Watch Extended" (PI: Gert G. Wagner, DFG-Grant Number WA 547/6-1).

**Abstract:** Academic data sharing is a way for researchers to collaborate and thereby meet the needs of an increasingly complex research landscape. It enables researchers to verify results and to pursue new research questions with "old" data. It is therefore not surprising that data sharing is advocated by funding agencies, journals, and researchers alike. We surveyed 2661 individual academic researchers across all disciplines on their dealings with data, their publication practices, and motives for sharing or withholding research data. The results for 1564 valid responses show that researchers across disciplines recognise the benefit of secondary research data for their own work and for scientific progress as a whole—still they only practice it in moderation. An explanation for this evidence could be an academic system that is not driven by monetary incentives, nor the desire for scientific progress, but by individual reputation—expressed in (high ranked journal) publications. We label this system a *Reputation Economy*. This special economy explains our findings that show that researchers have a nuanced idea how to provide adequate formal recognition for making data available to others—namely data citations. We conclude that data sharing will only be widely adopted among research professionals if sharing *pays* in form of reputation. Thus, policy measures that intend to foster research collaboration need to understand academia as a reputation economy. Successful measures must value intermediate products, such as research data, more highly than it is the case now.

# Introduction

Our modern research landscape today is characterized by a collaboration imperative (Bozeman & Boardman, 2014). Researchers specialize and a number of specialists need to be brought together to perform a noteworthy investigation. The most prominent form of research collaboration is the co-authored publication. It leads to more high-impact findings, and, in many fields, it is virtually impossible for a single investigator to develop meaningful insights (Wuchty et al. 2007). Besides co-authorship, academic data sharing is promising evolving form of research collaboration. Here, researchers make their primary datasets available to others. This allows for the verification of results and the reuse of existing datasets for new research questions. Open access to research data is therefore a useful measure for more transparency and collaboration in academic research. It is thus hardly surprising that Neelie Kroes, the European Commissioner for Digital Agenda, predicts that open data in research *"will boost Europe's innovation capacity and give citizens quicker access to the benefits of scientific discoveries"* (Kroes, 2012).

Despite the fact that open research data is supported by funding agencies for academic research (e.g., Kroes, 2013; NIH, 2014; Bijsterbosch, 2013), by an increasing number of academic journals, and even by many researchers (e.g., Tenopir, 2011) open research data is rarely practised by researchers themselves. It remains to a large extent an ideal that is rarely implemented (see for example Andreoli-Versbach & Mueller-Langer, 2014).

In order to explain the discrepancy between the expected benefit for scientific progress and the individual researcher's behaviour, we conducted an online survey among researchers in Germany and abroad. We were able to analyse the answers of 1564 researchers across all areas of academic research. They show how researchers handle data and what explains their data withholding strategies. We conclude that academia is a *reputation economy*, an exchange system that is driven by individual reputation beyond money and status. In this regard, data sharing will only see widespread adoption among research professionals if it *pays* in the form of reputation. In this study, we aim to provide empirical evidence for science policies and research data infrastructure that meet the individual researchers' demands.

# Background

Over the last decade, there has been a steep increase in interest in the topic of academic data sharing. Fecher et al. (2014a) conducted a systematic review on what drives data sharing in academia that is summarized in the following.

Tenopir et al. (2011) found that the willingness to make data available increases with *age*: Researchers older than 50 show more interest than their younger counterparts. This behaviour is explained by the higher degree of competition for tenure and professional success that younger researchers face. The authors further point out that researchers that have teaching obligations (i.e., less time to conduct research) are more reluctant to share. Several authors (Eschenfelder & Johnson, 2011, Haddow et al., 2011, Jarnevich et al., 2007, Pearce & Smith, 2011) highlight the importance of *control*—that is, who gets access to the data when. Van Horn and Gazzaniga (2013), for instance, explain that an embargo period, allowing data collectors to publish articles with the data first, would enable data sharing. Several authors refer to the effort of making data available as a key barrier (e.g., Acord and Harley, 2012, Enke et al., 2011, Gardner et al., 2003) or the time needed to help others to make sense of the shared data (Wallis et al., 2013).

*Recognition* as in *"what's in it for me?"* is also frequently mentioned in the literature. The consensus among many authors is that publishing data does not receive sufficient recognition. Making research data available to others will in this regard only see widespread adoption if data providers feel recognised and rewarded (Gardner et al., 2003; Ostell, 2009; Andreoli-Versbach & Mueller-Langer, 2014). Providing data should be treated similarly to academic authorship in publications (German Data Forum, 2011). K*nowledge* plays an important role in the data sharing behaviour of academic researchers. A lack of knowledge regarding curating and archiving datasets, for instance, prevents researchers from sharing (Haendel et al., 2012; Van Horn and Gazzaniga, 2013; Rendtel, 2011). Some scholars (Piwowar et al., 2008; Noor et al., 2008) therefore argue that data sharing needs to be integrated in the curriculum in data intensive fields of study.

A concern researchers have in regard to the reuse of their data is *adverse use*. Adverse use comprises any form of behaviour by the secondary data user that results in an undesirable outcome for the person that made the data available, for instance falsification. While falsification of research results is desirable for the research system as a whole it potentially hinders individual researchers from sharing (Costello, 2009; Acord and Harley, 2012; Pearce and Smith, 2011). Further, the fear of competitive misuse is an argument that can be found throughout literature on the topic (Acord and Harley, 2012; Dalgleish et al., 2012; Masum et al., 2013; Noor et al., 2008). Researchers, the argument goes, are hesitant to share data as they fear that someone else might publish an idea that they themselves want to publish. Some authors point out that the fear of commercial misuse can prevent researchers from making their data available (e.g., Gardner et al., 2003). Also, the possibility of a flawed interpretation is mentioned as a reason not to share research data (Cooper, 2007; Enke et al., 2011; Perrino et al., 2013). This could for instance mean that the re-user does not understand measures or populations. Lastly, the re-user's organization might influence researchers not to share data. This may be due to fears of inadequate containment and data protection or because they wish to avoid commercial use (Fernandez et al., 2012; Tenopir et al., 2011).

# Methodology

The aim of our survey is to give an empirically derived explanation for the discrepancy between the expected benefits of open research data and the individual researcher's data withholding strategy. The survey presented in this article is based on a systematic review on the topic and some results of a survey among secondary data users of the German Socio-Economic Panel (Hebing et al., 2014a; Wagner et al., 2007). Figure 1 illustrates the overall research design.

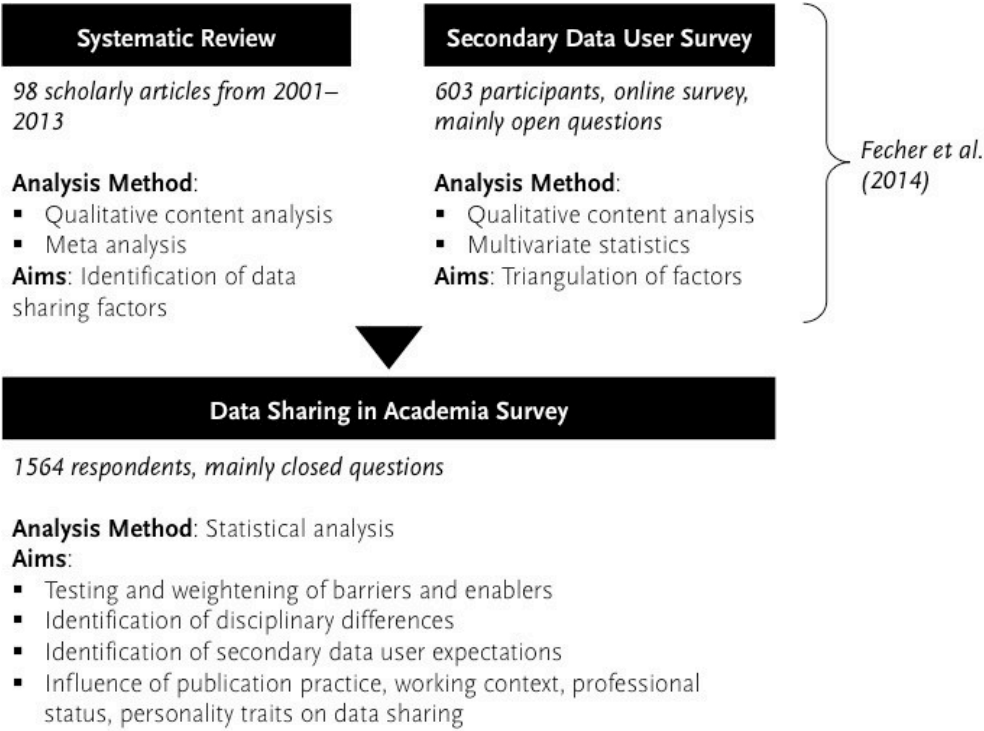

*Figure 1: Research Framework*

**Survey Instrument**

The survey contains mainly closed multiple-choice questions, rating scales and a few open questions. It covers questions on sociodemographics, the individual working context, the publication culture, the influence of control over data, common impediments and incentives for sharing data, expectations for using secondary data and personality traits. The questionnaire and a detailed description can be found on the project's Github page (Fecher et al., 2014b).

In this article we analyse the answers on publication culture, influence of control over data, common impediments and incentives for sharing data. The results of the latter two questions, regarding expectations for using secondary data and personality traits, will be covered in different articles.

When designing a survey, it is important to diminish the perceived intrusiveness, fear of disclosure and social desirability. Self-administration through the privacy of an online survey has been shown to decrease social desirability-biases—compared to personal interviews (Tourangeau and Yan, 2007). We assured the respondents confidentiality in the invitation to the survey as well as at the beginning of the survey, where we linked to the data privacy policies of the German Institute for Economic Research (DIW Berlin). Before handing out the survey, we conducted two pretesting rounds; the first with researchers from data-intensive fields on the usability and comprehensibility of the survey, the second with experts on data archiving and data reuse on the questions.

We conducted the online survey from October to November 2014. It was administered online via LimeSurvey. We contacted the faculty heads ("deans") of 20 large, medium-sized and small universities and universities of applied sciences (as measured by number of students) in Germany. Additionally we contacted the scientific directors of the four biggest German research organizations, the Max Planck Society, the Leibniz Association, the Helmholtz Association and the Fraunhofer Gesellschaft. Furthermore we uploaded a link to our survey on the project website (http://www.leibniz-science20.de/en/) and on the website of the German Data Forum (www.ratswd.de). Additionally we sent the survey to several mailing lists, including our institutional mailing lists (e.g., Leibniz institutes, Global Network of Interdisciplinary Internet & Society, Humboldt Institute for Internet and Society). In the emails to the faculty leaders and in the introductory text of the survey, we specifically addressed researchers that work with data. Due to the distribution strategy, our sample is a convenient sample, which means that it is not representative for the entire population of academic researchers in Germany (nor worldwide).

## Sample

Overall, 2661 people started the questionnaire, but not all respondents finished it. We chose to exclude respondents who failed to answer any questions about their status, employer and their discipline. We also excluded respondents that had less than 20% of the questions answered. We were left with 1564 valid entries – which are about 59% of all respondents.

88% of those respondents work in German institutions. 12% of the respondents work in other countries. Although we contacted German institutions, the number of researchers outside Germany who responded to the survey was relatively high; this can be explained by respondents reached via the mailing lists and the website postings. The average age of the respondents is 38 years. Figure 2 shows the composition of our sample by academic status and disciplinary background of the respondents.

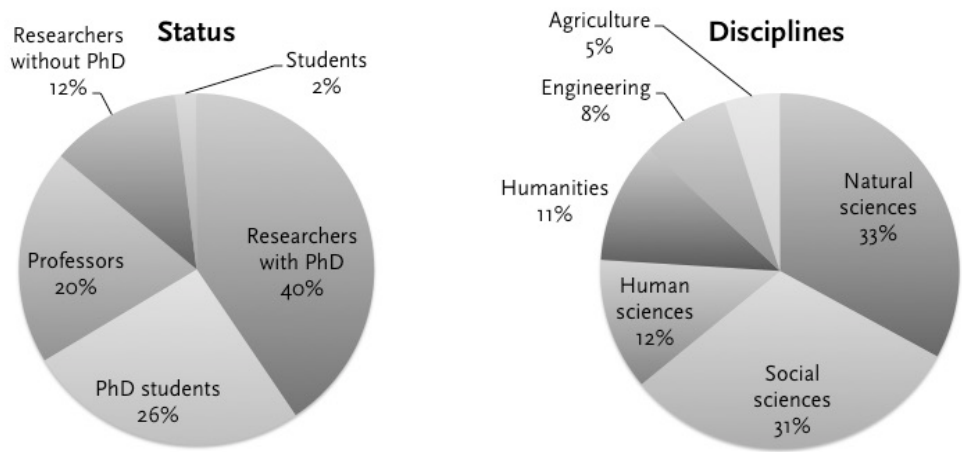

*Figure 2: Sample Composition*

Within the disciplines there are noteworthy differences in the gender distribution (see table 1). For example, of the 431 researchers from the natural sciences, the largest discipline group in our sample, 272 (63%) are male and 159 (37%) female—a fact that is important to understand gender differences in the sharing behaviour and attitude.

# Results

In the following we will first present the findings of the descriptive analysis of the survey. Subsequently we will present results that include the regression analysis for selected variables.

## Descriptive Results

In this section we present the descriptive results of the survey. This includes opinions on data sharing and the sharing culture, the materials a researcher would share as well as common impediments and motivators. We operationalize approval by combining the two most approving points on the 5–point Likert scale and rejection by combining the two most rejecting points. For instance, when we state that 74% of the respondents agree with a statement, we summarize those respondents that ticked 4 and 5 (towards *strongly agree*) on the Likert scale. The tables at the end of this document include the mean and standard deviation for each question as well as the percentage for each point in the rating scales.

### *Opinions on Sharing*

We asked about the general perception and the individual implementation among the researchers in our sample with a group of questions on opinions on data sharing and the individual sharing experience (table 2).

Most respondents (76%) agree that other researchers should publish their data. To the statement if making their own data available brings more disadvantages than advantages, 62% of the respondents disagree. 83% state, that making data available to other researchers benefits scientific progress. We asked whether it is common in their respective discipline/research community to make data available. Across all disciplines, 35% of the researchers say that it is common. More respondents (37%) state that it is uncommon to share data in their community. Within the single disciplines, the highest approval of this question can be found in natural science. 48% of researchers from natural science say it is common in their community to share data, followed by 42% of researchers from the human sciences. The numbers in social sciences (26%), in engineering (26%) and in the humanities (23%) are considerably lower (table 3).

When respondents were asked whether they themselves have shared research data in the past and with whom, 58% said they have shared with researchers they know personally, 49% shared with researchers within their respective institute or organization, 40% with researchers that work on similar topics and 13% publicly (see table 5). Across all disciplines, researchers widely agree that it is beneficial for scientific progress to make research data available to others (see table 2). Most researchers (88%) would use secondary data for pursuing their own original research question. 47% of the respondents would use secondary data to replicate or verify results (table 2).

*Impediments and Motivators*

We tested common impediments and motivators regarding making data available. By a wide margin, the main barrier was that *"other researchers could publish before me"* (80%); the test variable *"to publish before sharing"* (78%) is the second highest motivator for making data available to others (see table 4).

The second biggest impediment to sharing is if it was a *"major effort to share"* at 59%. At 46%, the concern that data could be misinterpreted ranks third. Few researchers (12%) are concerned about being *"criticized or falsified"*. The majority of the respondents (72%) instead disagrees that criticism or falsification would prevent them from making data available. 45% say that the concern that the data could be misinterpreted prevents them from sharing data.

Across disciplines, 79% of the respondents say that data citation would motivate them to make data available to others; only 9% say it would not. Despite the demand for more formal recognition, only 34% regard *"Co-authorship"* as a motivator. 44% of the researchers in our sample reject it. A majority of 65% of the respondents rejects financial support—only 17% see it as a motivator. It is noteworthy that the ranking for both the impediments and motivators for sharing research data is nearly identical across all disciplines (see figure 3).

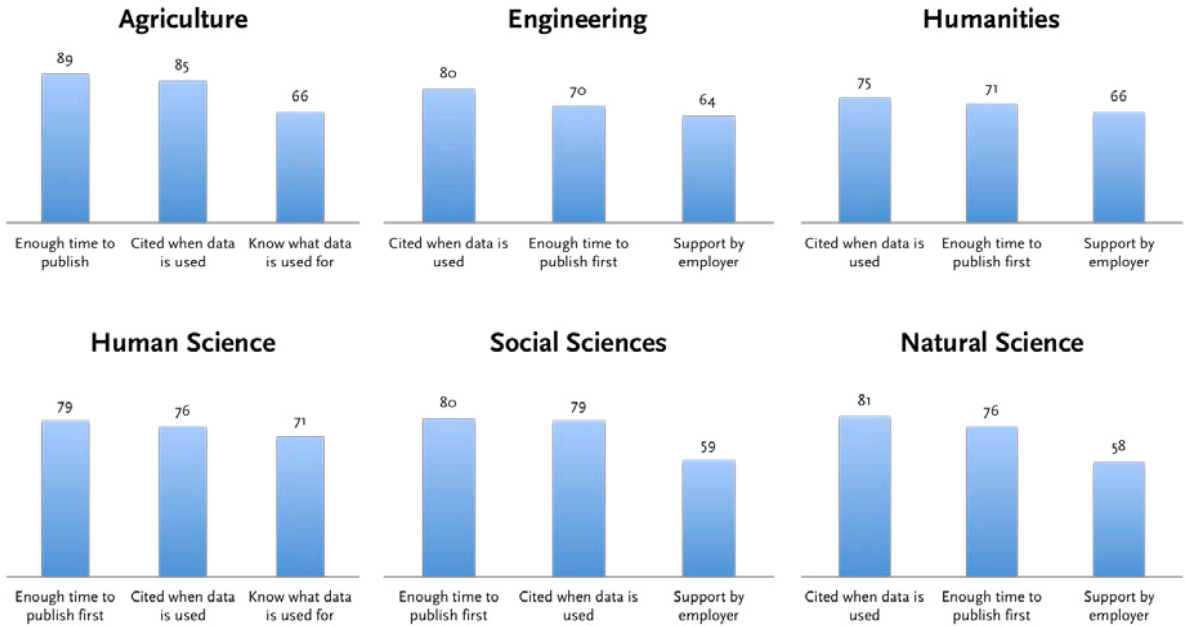

Figure 3: Top 3 Motivators by discipline

*Conditional Sharing*

In the survey we asked with whom researchers would be willing to share data and, using the same response categories (see table 5), whether they have already made data available to others (see table 5). To the question *"Who would you be willing to share your data with?"* 72% of the respondents replied that they would share with researchers whom they know personally. Only 32% said they would be willing to share their data publicly. 56% stated that they would make their data available to all non-commercial researchers. To the question whether and with whom they actually shared data, 58% said they shared data with researchers they know personally. 14% said they have made their data available to all non-commercial researchers. 13% have shared their data publicly in the past. *"Sharing with commercial researchers"* is rejected by the respondents. Only 16% would be willing to share data with commercial researchers, while 6% have shared with commercial researchers in the past. 16% said they have never shared data in the past. To the statement *"I would only share my data if I knew what the data was going to be used for"*, 50% of the researchers agree, 32% disagree (table 5). To the question concerning the conditions under which they would share data (see table 4) less than 1% answered *"Not at all"*. 35% would share on demand, 42% with a specific use agreement. 23% would share unconditionally. Most researchers would make data available if they could decide on the scope and modalities of the data reuse: *who* can access *what* kind of data *how* and *when*.

We asked researchers which research materials they would be willing to share with other researchers (table 2). We defined research material as including raw data, survey design, prepared data, data documentation, analysis scripts and code/software. Researchers are most likely to share their survey design (81%), followed by data documentation (78%); raw data ranks last (58%).

## Multivariate Analyses

We ran two multivariate regression models. The models use two dependent variables *"willingness to share"* and *"has shared data in past"*. We recoded the item batteries *"Who would you be willing to share your data with?"* and *"Have you ever shared your research data with others?"* into two binary variables, combining the categories 4 and 6 into "*willingness to share with a broader audience*" respectively "*has shared with a broader audience*". A 'broader audience' is hereby defined as the public or non-commercial researchers. As independent variables, we plugged in sociodemographic factors, the knowledge regarding data sharing, the professional status and the organizational context, the publication practice as well as the experience with secondary data. In some cases where we believe it contributes to a better understanding of the results, we included descriptive results.

### *Sociodemographic Factors*
The results of the regression analyses show that there is no significant influence of a researcher's age on his/hers willingness to make data available and if a respondent shared in the past (table 6). However, in a second logit model (without status), age did have a positive impact on the willingness to share data (table 7). Experience in academic research has no significant influence on the willingness to make data available. Female researcher in our sample are significantly less willing to make data available to other researchers than their male colleagues and significantly less likely to have shared data in the past (table 6). 9% of the female researchers have shared research data publicly in the past compared to 18% of the male respondents.

### *Knowledge*

In order to identify disciplinary differences in the knowledge regarding data reuse, we inquired if researchers know a) where and how they can find relevant secondary data for their work and b) where and how to make research data available to others. Overall 57% of respondents said that they know where to find secondary data and 50% know where and how to publish research data. Knowledge varies between disciplines. Natural scientists know significantly more about where and how they can publish data than respondents from other disciplines (see table 8). Researchers from engineering and agriculture know significantly less about where they can find secondary data for their research (see table 9) than respondents from other disciplines. Knowledge about how to publish data increases significantly with age (table 8). Researchers that *know how* to make data available to others are significantly more willing to make data available. They are also more likely to have shared data in the past (table 6). Researchers that used secondary data before are significantly more willing to share data.

### *Professional Status and Organizational Context*

Our findings show no effect of the professional status on the willingness to make data available and if a researcher has shared in the past (see table 6). We tested if it makes a difference if a researcher is employed at a university, a university of applied sciences or an institute. Our results show no effect of the organizational context on the willingness to share or the likelihood that a researcher has shared in the past.

### *Publishing Preferences*

Regarding the importance of article publications for academic success and professional development, we tested how publication preferences influence data sharing. 79% of respondents agreed that when publishing results "*reputation/impact*" is important, which is thereby the most important criteria when publishing results, followed by "*fast publishing process*" with 52% agreement and lastly "Open Access" with agreement of 39% (table 11). Asked if it would deter researchers from sharing data if a journal required the publication of research data, most respondents disagree (74%) with little variation (see table 2). Compared by discipline, the biggest emphasis on reputation can be found in the agricultural sciences with 84%, while the lowest was in the humanities with 67%. Open access has relatively high approval among natural scientists (47%) and humanities scholars (47%). 40% of the respondents from human sciences, 33% from engineering and 30% from social science agree that Open Access is an important publication criteria. Fast publishing is similarly important in all disciplines—it is lowest in the social sciences with 50% and highest in the agricultural sciences with 57% (table 10).

Researchers that value Open Access highly are significantly more likely to make their data available. 31% of the respondents who stated that Open Access is important are willing to share their data unconditionally whereas only 22% of those who value reputation/impact highly are willing to share their data unconditionally (table 6). It is noteworthy that those who favour a "*fast publishing process*" are significantly less willing to make their data available than those who do not.

## Discussion

Academic data sharing is an accepted form of research collaboration and perceived as beneficial for scientific progress among academic researchers. The innovative capacity of shared research data is evident for researchers across all disciplines. They use secondary data themselves to pose new research questions and to verify others results. Furthermore, they think that researchers should generally make data available. The ideal of making the research system as a whole better by sharing data is understood and supported by most researchers. However this understanding does not translate into action. Hardly any of the respondents in our survey has shared data publicly. However, many researchers say they would be willing to share data. This discrepancy between the willingness to share and the individual researcher's behaviour in view of its importance for scientific progress points to an untapped potential in academic research.

Our empirical results suggest that the improvable state of academic data sharing is in many respects a function of reputation considerations—a rationale that holds true across all disciplines. Reputation/impact is the most important factor when publishing research. The greatest concern about data sharing is that other researchers could publish before the researcher in question; to publish first is the most important sharing condition. This indicates that in the researchers perception, data publications are valued far less for a researcher's reputation building than research articles. At the same time, researchers that are motivated strongly by fast publication speeds are less likely to publish research data as they are even more motivated to work on new publications. We see a stronger impetus to share research data among those researchers that also make their publications freely available (i.e., publishing Open Access). Making data publicly available at an early stage is in this regard also a question of attitude. The latter is in line with previous findings (Andreoli-Versbach & Mueller-Langer 2014). Hardly anyone is motivated to share his/her data for financial incentives; instead, the respondents strongly reject financial compensation and reject to make data available to commercial researchers. The results further show that researchers are generally not inhibited by mandatory data sharing policies and not afraid to be falsified or criticized by re-analyses of their data. The core impediment to making data available is the lack of formal recognition of this task. This results in data withholding strategies and the decision to opt for traditional publications. Researchers have a nuanced idea of adequate recognition, as they do not demand co-authorship, which could be regarded as non-ethical in some disciplines, but data citations for making data available.

Reputation, not money or a common good orientation, seems to foster the exchange of information among researchers; a system that we label 'reputation economy'. In this regard we suggest rethinking reputation and impact considerations in academic publishing in favour of research intermediaries, in particular data. To develop a stronger culture of data sharing, policy makers, funding agencies and research organizations need to value the archiving and reusing of data more highly and, considering the effort of producing and archiving good data, "*give credit where credit is due*" (German Data Forum, 2011, p.38 ). In this regard, secondary data needs to be discoverable, citable and trackable. The use of secondary data needs to have a positive impact on the professional advancement and academic standing of the data provider.

Making research data available to others is a prime example of openness in research as it fosters collaboration and transparency. Open access to research data means the archiving and release of data in a digitally processable form for re-use (almost) from the point of collection (e.g., Carlson & Anderson, 2007)—an ideal that that cannot be reconciled with empirical reality. In this regard, our results also demonstrate the limits of openness from the individual researcher's perspective.

# Tables

| Table 1: Discipline Gender ||
|---|---|
| B: Discipline vs. Gender (in %) | Item-Responserate: 83% |

|  | **Male** | **Female** |
|---|---|---|
| **Natural science** | 63,11 | 36,89 |
| **Social science** | 52,88 | 47,12 |
| **Human science** | 43,05 | 56,95 |
| **Engineering** | 77,32 | 22,68 |
| **Humanities** | 53,47 | 46,53 |
| **Agricultural science** | 52,31 | 47,69 |
| *Total* | *56,98* | *43,02* |

# Tables

## Table 2: Descriptive Results

| Opinions on data sharing | | | | | | Responses in percent: | | | | |
|---|---|---|---|---|---|---|---|---|---|---|
| *On a scale from 1 to 5; 1="Strongly disagree" - 5="Agree completely"* | Obs. | IRR* (%) | Mean | Std. Dev. | 95% Conf. Interval | Strongly Disagree | 2 | 3 | 4 | Agree Completely |
| **Researchers should generally publish their data** | 1491 | 95,33 | 4,10 | 1,00 | 4,05  4,15 | 1,95 | 5,90 | 16,57 | 31,32 | 44,27 |
| **I have more disadvantages than advantages when I share my data with others.** | 1419 | 90,73 | 2,32 | 1,16 | 2,26  2,38 | 28,61 | 33,19 | 21,35 | 11,49 | 5,36 |
| **It is common in my discipline / research community to share data.** | 1436 | 91,82 | 2,95 | 1,21 | 2,89  3,02 | 13,86 | 23,33 | 27,37 | 24,44 | 11,00 |
| **I know where and how I can find relevant data for my work.** | 1449 | 92,65 | 3,49 | 1,13 | 3,44  3,55 | 5,66 | 15,25 | 22,43 | 37,34 | 19,32 |
| **I know where and how I can make data that I collected available to others.** | 1466 | 93,73 | 3,35 | 1,28 | 3,28  3,42 | 9,48 | 19,51 | 20,60 | 27,35 | 23,06 |
| **It deters me from publishing when a journal requires the publication of my data.** | 1412 | 90,28 | 1,94 | 1,14 | 1,88  2,00 | 48,30 | 25,71 | 13,95 | 7,86 | 4,18 |
| **Freely available research data is a great contribution to scientific progress** | 1449 | 92,65 | 4,34 | 0,94 | 4,29  4,39 | 1,73 | 3,80 | 11,32 | 25,05 | 58,11 |
| **Which materials would you be willing to share with other researchers?** | | | | | | Responses in percent: | | | | |
| *On a scale from 1 to 5; 1="Does not apply at all" - 5="Applies completely"* | Obs. | IRR* (%) | Mean | Std. Dev. | 95% Conf. Interval | Strongly Disagree | 2 | 3 | 4 | Agree Completely |
| **Raw data** | 1352 | 86,45 | 3,51 | 1,36 | 3,44  3,58 | 11,02 | 15,83 | 15,38 | 26,85 | 30,92 |
| **Survey design** | 1306 | 83,50 | 4,20 | 1,03 | 4,14  4,25 | 3,22 | 4,82 | 11,03 | 30,86 | 50,08 |
| **Prepared data** | 1373 | 87,79 | 3,99 | 1,06 | 3,93  4,05 | 2,77 | 8,01 | 16,10 | 33,72 | 39,40 |
| **Data documentation** | 1334 | 85,29 | 4,17 | 0,97 | 4,12  4,22 | 1,95 | 4,20 | 15,67 | 31,48 | 46,70 |
| **Analysis scipts** | 1226 | 78,39 | 3,58 | 1,25 | 3,51  3,65 | 6,69 | 15,91 | 20,23 | 27,00 | 30,18 |
| **Code/software** | 1198 | 76,60 | 3,56 | 1,30 | 3,48  3,63 | 8,43 | 15,36 | 19,20 | 26,21 | 30,80 |
| **What would u like to do with secondary data?** | | | | | | Responses in percent: | | | | |
| *On a scale from 1 to 5; 1="Does not apply at all" -* | Obs. | IRR* (%) | Mean | Std. Dev. | 95% Conf. Interval | Strongly Disagree | 2 | 3 | 4 | Agree Completely |

| 5="Applies completely" | | | | | | | | | | |
|---|---|---|---|---|---|---|---|---|---|---|
| Use them for (my own) original research questions | 1287 | 82,29 | 4,41 | 0,86 | 4,36 | 4,45 | 1,63 | 2,49 | 7,61 | 30,23 | 58,04 |
| Use them to replicate and verify research findings | 1278 | 81,71 | 3,27 | 1,34 | 3,20 | 3,34 | 11,42 | 21,36 | 20,19 | 22,93 | 24,1 |

*IRR= Item-Response-Rate

| Table 3: Sharing is common in my discipline |||||||||||
|---|---|---|---|---|---|---|---|---|---|---|
| It is common in my discipline/research community to share data. ||||| Responses in percent: |||| Item-responserate: 90.5% ||
| On a scale from 1 to 5; 1="Strongly Disagree" - 5="Agree Completely" | Obs. | Mean | Std. Dev. | 95% Conf. Interval || Strongly Disagree | 2 | 3 | 4 | Agree Completely |
| Natural science | 472 | 3,35 | 1,16 | 3,24 | 3,45 | 6,36 | 18,64 | 26,69 | 30,72 | 17,58 |
| Social science | 425 | 2,66 | 1,19 | 2,55 | 2,77 | 19,76 | 27,29 | 26,59 | 19,76 | 6,59 |
| Human sciance | 168 | 3,15 | 1,16 | 2,97 | 3,33 | 10,71 | 17,26 | 29,76 | 30,95 | 11,31 |
| Engineering | 119 | 2,89 | 1,07 | 2,70 | 3,09 | 10,92 | 22,69 | 40,34 | 18,49 | 7,56 |
| Humanities | 168 | 2,48 | 1,24 | 2,29 | 2,67 | 26,19 | 29,76 | 20,83 | 16,07 | 7,14 |
| Agricultural science | 64 | 3,00 | 1,10 | 2,73 | 3,27 | 7,81 | 28,13 | 28,13 | 28,13 | 7,81 |
| Total | 1416 | 2,96 | 1,21 | 2,90 | 3,02 | 13,70 | 23,16 | 27,54 | 24,58 | 11,02 |

| Table 4: Enablers and Barriers for Data Sharing |||||||||||
|---|---|---|---|---|---|---|---|---|---|---|
| **Enablers: I would only share my data…** |||||| Responses in percent: |||||
| *On a scale from 1 to 5; 1="Does not apply at all" - 5="Applies completely"* | Obs. | IRR* (%) | Mean | Std. Dev. | 95% Conf. Interval || **Strongly Disagree** | 2 | 3 | 4 | **Agree Completely** |
| if I knew what the data were going to be used for. | 1420 | 90,79 | 3,27 | 1,38 | 3,20 | 3,34 | 14,51 | 17,82 | 18,03 | 25,56 | 24,08 |
| if sharing the data enabled me to get into contact with other researchers. | 1430 | 91,43 | 3,04 | 1,24 | 2,97 | 3,10 | 14,62 | 18,74 | 26,99 | 27,55 | 12,10 |
| if I had enough time beforehand, to publish on the basis of my data. | 1420 | 90,79 | 4,10 | 1,09 | 4,04 | 4,16 | 4,08 | 5,85 | 12,54 | 31,06 | 46,48 |
| if I knew who would be able to access the data. | 1424 | 91,05 | 3,18 | 1,39 | 3,11 | 3,25 | 16,36 | 17,56 | 19,59 | 24,51 | 21,98 |
| if my employer supported me actively (e.g. by providing technical support, time). | 1389 | 88,81 | 3,59 | 1,25 | 3,52 | 3,65 | 8,50 | 12,10 | 19,80 | 31,61 | 28,01 |
| if I were cited in publications using my data. | 1420 | 90,79 | 4,21 | 1,08 | 4,15 | 4,27 | 3,52 | 5,99 | 11,20 | 24,65 | 54,65 |
| if I were given a co-authorship of articles using my data. | 1414 | 90,41 | 2,86 | 1,42 | 2,79 | 2,94 | 22,49 | 21,85 | 21,43 | 15,42 | 18,81 |
| if I received financial compensation for the effort. | 1393 | 89,07 | 2,19 | 1,28 | 2,12 | 2,26 | 40,92 | 24,05 | 17,95 | 9,12 | 7,97 |
| *IRR= Item Response Rate |||||||||||
| **Barriers: I would not share my data…** |||||| Responses in percent: |||||
| *On a scale from 1 to 5;1="Does not apply at all" - 5="Applies completely"* | Obs. | IRR* (%) | Mean | Std. Dev. | 95% Conf. Interval || **Strongly Disagree** | 2 | 3 | 4 | **Agree Completely** |
| if other researchers could use my data to publish before me. | 1409 | 90,09 | 4,25 | 1,17 | 4,19 | 4,31 | 5,61 | 5,75 | 8,73 | 18,03 | 61,89 |
| if others could criticize or falsify my work. | 1402 | 89,64 | 1,99 | 1,18 | 1,93 | 2,05 | 46,65 | 25,89 | 15,05 | 6,70 | 5,71 |
| if the data could be misinterpreted. | 1383 | 88,43 | 3,16 | 1,35 | 3,08 | 3,23 | 15,84 | 17,43 | 21,04 | 26,75 | 18,94 |
| if the data collection required considerable effort. | 1392 | 89,00 | 2,57 | 1,26 | 2,50 | 2,63 | 26,51 | 23,42 | 24,71 | 17,53 | 7,83 |
| if a major effort were required to share the data. | 1403 | 89,71 | 3,55 | 1,19 | 3,49 | 3,61 | 7,84 | 11,69 | 21,53 | 35,64 | 23,31 |
| *IRR= Item Response Rate |||||||||||

| Table 5: Conditional Sharing ||||||||
|---|---|---|---|---|---|---|---|
| **Who would you be willing to share your data with?** ||||||| Responses in percent: |
| *binary, multiple choice; 1= selected 0= not selected* | Obs. | Mean | Std. Error | 95% Conf. Interval || **Selected** | **Not selected** |
| With researchers who I know personally | 1564 | 0,72 | 0,011 | 0,70 | 0,74 | 72,19 | 27,81 |
| With researchers within my research institute or my organisation | 1564 | 0,68 | 0,012 | 0,65 | 0,70 | 67,65 | 32,35 |
| With researchers who work on similar topics | 1564 | 0,70 | 0,012 | 0,68 | 0,73 | 70,46 | 29,54 |
| With all non-commercial researchers | 1564 | 0,56 | 0,013 | 0,54 | 0,58 | 56,01 | 43,99 |
| With commercial researchers | 1564 | 0,16 | 0,009 | 0,14 | 0,18 | 16,24 | 83,76 |
| With the public | 1564 | 0,32 | 0,012 | 0,29 | 0,34 | 31,52 | 68,48 |
| **Have you ever shared your research data with others?** ||||||| Responses in percent: |
| *binary, multiple choice; 1= selected 0= not selected* | Obs. | Mean | Std. Error | 95% Conf. Interval || **Selected** | **Not selected** |
| Yes, with researchers who I know personally | 1564 | 0,58 | 0,012 | 0,56 | 0,61 | 58,06 | 41,94 |
| Yes, with researchers within my institute or my organisation | 1564 | 0,49 | 0,013 | 0,47 | 0,52 | 49,36 | 50,64 |
| Yes, with researchers who work on similar topics | 1564 | 0,40 | 0,012 | 0,38 | 0,43 | 40,35 | 59,65 |
| Yes, with all non-commercial researchers | 1564 | 0,14 | 0,009 | 0,13 | 0,16 | 14,45 | 85,55 |
| With commercial researchers | 1564 | 0,06 | 0,006 | 0,05 | 0,07 | 5,88 | 94,12 |
| Yes, with the public | 1564 | 0,13 | 0,009 | 0,11 | 0,15 | 13,11 | 86,89 |
| No | 1564 | 0,16 | 0,009 | 0,14 | 0,18 | 16,30 | 83,70 |

| Table 6: Complete Logit-Regression | | |
|---|---|---|
| | A | B |
| Logit regression (Odds-Ratios) | Willingness to share | Shared data in past |
| **Age centered** | 0.983 | 0.997 |
| | (0.0103) | (0.0113) |
| **Sex (Ref.: male)** | 0.528*** | 0.510*** |
| | (0.0857) | (0.102) |
| **Employer (Ref.: Institutes)** | 0.994 | 0.994 |
| | (0.170) | (0.203) |
| **Status (Ref.: Prof)** | | |
| Student | 0.603 | 0.152* |
| | (0.382) | (0.173) |
| PhD Student | 1.151 | 0.491* |
| | (0.370) | (0.182) |
| Researcher without PhD | 1.874* | 0.763 |
| | (0.695) | (0.307) |
| Researcher with PhD | 1.118 | 1.037 |
| | (0.270) | (0.274) |
| **Natural Science** | 1.131 | 1.106 |
| (Ref.: all other disciplines) | (0.210) | (0.230) |
| **Importance: Open Access** | 1.238*** | 1.182** |
| | (0.0771) | (0.0865) |
| **Importance: Reputation/Impact** | 1.025 | 0.876 |
| | (0.0804) | (0.0752) |
| **Importance: fast publishing** | 0.793*** | 0.914 |
| | (0.0636) | (0.0796) |
| **Sensitive Data** | 0.984 | 1.022 |
| | (0.0546) | (0.0672) |
| **qualtoquant** | 0.963 | 1.011 |
| | (0.0333) | (0.0431) |
| **Uses secondary data** | 1.421** | 2.781*** |
| | (0.253) | (0.724) |
| **Wants to use secondary data for...** | | |
| ...original research | 1.412*** | 1.148 |
| | (0.133) | (0.132) |
| ...replication | 0.982 | 1.064 |
| | (0.0573) | (0.0735) |
| **Knows where and how to...** | | |
| ...publish data | 1.277*** | 2.107*** |
| | (0.0906) | (0.197) |
| ...find data | 0.865* | 1.084 |
| | (0.0679) | (0.102) |
| Constant | 0.435 | 0.00529*** |
| | (0.309) | (0.00464) |
| Observations | 927 | 927 |
| *Pseudo-R2* | *0.0729* | *0.206* |
| *LR-Test* | *81.85* | *209.9* |
| *df* | *18* | *18* |
| Standard error in parentheses | | *** p<0.01, ** p<0.05, * p<0.1 |

| Table 7: Logit Regression without Status | | |
|---|---|---|
| | **C** | **D** |
| Logit Regression (Odds-Ratios) | **willingness to share** | **has shared data in past** |
| | | |
| **Age centered** | 0.982** | 1.011 |
| | (0.00756) | (0.00849) |
| **Status not included in model** | | |
| **All other Variables are the same as in Table 5** | | |
| Constant | 0.414 | 0.00448*** |
| | (0.267) | (0.00364) |
| Observations | 956 | 956 |
| *Pseudo-R2* | *0.0689* | *0.194* |
| *LR-Test* | *79.02* | *206.1* |
| *df* | *14* | *14* |
| *** p<0.01, ** p<0.05, * p<0.1 | | Standard error in parentheses |

| Table 8: Linear Regression - Knowledge | | |
|---|---|---|
| | 1 | 2 |
| Linear Regression: | **Knowledge: finding Data** | **Knowledge: publishing data** |
| **Age centered** | 0.00847* | 0.0130** |
| | (0.00488) | (0.00542) |
| **Sex (Ref.: male)** | 0.0879 | -0.0782 |
| | (0.0763) | (0.0848) |
| **Employer (Ref.: Institutes)** | -0.0909 | -0.0161 |
| | (0.0799) | (0.0888) |
| **Status (Ref.: Prof)** | | |
| Student | 0.0636 | -0.755** |
| | (0.298) | (0.332) |
| PhD Student | -0.183 | -0.594*** |
| | (0.150) | (0.166) |
| Researcher without PhD | 0.109 | -0.421** |
| | (0.165) | (0.183) |
| Researcher with PhD | -0.0361 | -0.182 |
| | (0.115) | (0.128) |
| **Natural Science** | -0.0595 | 0.271*** |
| (Referenz: all others) | (0.0855) | (0.0951) |
| **Importance: Open Access** | -0.0145 | 0.0979*** |
| | (0.0288) | (0.0320) |
| **Importance: Reputation/Impact** | 0.0478 | 0.0491 |
| | (0.0362) | (0.0403) |
| **Importance: fast publishing** | 0.0450 | 0.0231 |
| | (0.0360) | (0.0400) |
| **Sensitive Data** | -0.0113 | -0.0261 |
| | (0.0261) | (0.0290) |
| **qualtoquant** | -0.000956 | 0.0321* |
| | (0.0161) | (0.0179) |
| **Uses secondary data** | 0.693*** | 0.457*** |
| | (0.0828) | (0.0920) |
| **Wants to use secondary data for…** | | |
| …original research | 0.0397 | 0.0322 |
| | (0.0452) | (0.0503) |
| …replication | 0.0371 | 0.0799*** |
| | (0.0273) | (0.0303) |
| Constant | 2.505*** | 2.277*** |
| | (0.324) | (0.360) |
| | | |
| Observations | 927 | 927 |
| *R-squared* | *0.106* | *0.146* |

Standard error in parentheses      \*\*\* p<0.01, \*\* p<0.05, \* p<0.1

The dependent variable ranges from 1 to 5, from disagreement to agreement.

## Table 9: Knowledge and Disciplines

| Knowledge: finding data | | | | | | Responses in percent: | | | | Item-Responserate=91% |
|---|---|---|---|---|---|---|---|---|---|---|
| On a scale from 1 to 5; 1="Strongly disagree"- 5="Agree completely" | Obs. | Mean | Std. Dev. | [95% Conf. Intervall] | | Strongly Disagree | 2 | 3 | 4 | Agree Completely |
| Natural science | 461 | 3,54 | 1,09 | 3,44 | 3,64 | 4,77 | 13,67 | 22,99 | 40,13 | 18,44 |
| Social science | 446 | 3,58 | 1,15 | 3,47 | 3,69 | 5,16 | 14,35 | 21,52 | 35,2 | 23,77 |
| Human sciance | 169 | 3,47 | 1,16 | 3,3 | 3,65 | 6,51 | 17,75 | 14,2 | 44,97 | 16,57 |
| Engineering | 120 | 3,25 | 1,07 | 3,06 | 3,44 | 6,67 | 16,67 | 32,5 | 33,33 | 10,83 |
| Humanities | 164 | 3,42 | 1,24 | 3,23 | 3,61 | 7,93 | 17,68 | 21,95 | 29,27 | 23,17 |
| Agricultural science | 67 | 3,19 | 1,12 | 2,92 | 0 | 7,46 | 20,9 | 26,87 | 34,33 | 10,45 |
| *Total* | *1427* | *3,49* | *1,14* | *3,43* | *3,55* | *5,75* | *15,42* | *22,35* | *37,07* | *19,41* |

| Knowledge: publishing data | | | | | | Responses in percent: | | | | Item-Responserate=92% |
|---|---|---|---|---|---|---|---|---|---|---|
| On a scale from 1 to 5; 1="Strongly disagree"- 5="Agree completely" | Obs. | Mean | Std. Dev. | [95% Conf. Intervall] | | Strongly Disagree | 2 | 3 | 4 | Agree Completely |
| Natural science | 473 | 3,63 | 1,17 | 3,52 | 3,74 | 4,86 | 15,43 | 18,6 | 34,04 | 27,06 |
| Social science | 443 | 3,12 | 1,38 | 2,99 | 3,25 | 15,12 | 22,35 | 19,64 | 20,99 | 21,9 |
| Human sciance | 176 | 3,3 | 1,26 | 3,11 | 3,49 | 7,95 | 23,3 | 21,02 | 26,14 | 21,59 |
| Engineering | 118 | 3,34 | 1,1 | 3,14 | 3,54 | 4,24 | 21,19 | 25,42 | 34,75 | 14,41 |
| Humanities | 167 | 3,24 | 1,4 | 3,03 | 3,45 | 14,37 | 18,56 | 22,75 | 17,37 | 26,95 |
| Agricultural science | 67 | 3,27 | 1,14 | 2,99 | 3,55 | 7,46 | 17,91 | 28,36 | 32,84 | 13,43 |
| *Total* | *1444* | *3,35* | *1,29* | *3,28* | *3,41* | *9,56* | *19,46* | *20,71* | *27,15* | *23,13* |

## Table 10: Knowledge and Age

| Knowledge: publishing data | | | | | | Responses in percent: | | | | Item-Responserate=81% |
|---|---|---|---|---|---|---|---|---|---|---|
| On a scale from 1 to 5; 1="Strongly disagree"- 5="Agree completely" | Obs. | Mean | Std. Dev. | [95% Conf. Intervall] | | Strongly Disagree | 2 | 3 | 4 | Agree Completely |
| **Age: 22-34** | 594 | 3,01 | 1,269 | 2,91 | 3,11 | 13,3 | 25,93 | 21,55 | 25,08 | 14,14 |
| **Age: 35-47** | 409 | 3,56 | 1,23 | 3,44 | 3,67 | 6,85 | 14,91 | 21,76 | 28,85 | 27,63 |
| **Age: 48-60** | 214 | 3,74 | 1,232 | 3,57 | 3,9 | 5,61 | 14,02 | 16,36 | 28,97 | 35,05 |
| **Age: 60-71** | 53 | 3,81 | 1,194 | 3,49 | 4,13 | 5,66 | 11,32 | 13,21 | 35,85 | 33,96 |
| *Total:* | *1270* | *3,34* | *1,29* | *3,27* | *3,41* | *9,61* | *19,76* | *20,39* | *27,4* | *22,83* |

gamma = 0.2961 ASE = 0.031; Kendall's tau-b = 0.2118 ASE = 0.023

## Table 11: Publication Preferences by discipline

| Importance: Open Acces | | | | | | Responses in percent: | | | | Item-Responserate: 90% |
|---|---|---|---|---|---|---|---|---|---|---|
| | Obs. | Mean | Std. Dev. | [95% Conf. Interval] | | Not important | 2 | 3 | 4 | Very important |
| **Natural science** | 480 | 3,30 | 1,31 | 3,18 | 3,41 | 13,33 | 12,71 | 27,29 | 24,38 | 22,29 |
| **Social science** | 427 | 2,78 | 1,28 | 2,66 | 2,90 | 20,37 | 22,48 | 27,63 | 17,56 | 11,94 |
| **Human science** | 167 | 3,11 | 1,24 | 2,92 | 3,30 | 13,77 | 15,57 | 31,14 | 25,15 | 14,37 |
| **Engineering** | 111 | 2,87 | 1,32 | 2,63 | 3,12 | 19,82 | 19,82 | 27,03 | 19,82 | 13,51 |
| **Humanities** | 158 | 3,36 | 1,36 | 3,15 | 3,58 | 13,29 | 12,66 | 27,22 | 18,35 | 28,48 |
| **Agricultural science** | 65 | 3,14 | 1,16 | 2,85 | 3,43 | 10,77 | 13,85 | 40 | 21,54 | 13,85 |
| *Total* | *1408* | *3,08* | *1,31* | *3,02* | *3,15* | *15,91* | *16,62* | *28,41* | *21,24* | *17,83* |

| Importance: Reputation/Impact | | | | | | Responses in percent: | | | | Item-Responserate: 92% |
|---|---|---|---|---|---|---|---|---|---|---|
| | Obs. | Mean | Std. Dev. | [95% Conf. Interval] | | Not important | 2 | 3 | 4 | Very important |
| **Natural science** | 488 | 4,15 | 0,98 | 4,06 | 4,23 | 2,05 | 5,12 | 13,73 | 34,22 | 44,88 |
| **Social science** | 441 | 4,24 | 0,97 | 4,15 | 4,33 | 2,27 | 4,99 | 9,52 | 32,88 | 50,34 |
| **Human science** | 172 | 4,27 | 1,01 | 4,12 | 4,43 | 2,91 | 4,65 | 9,88 | 27,33 | 55,23 |
| **Engineering** | 116 | 4,08 | 1,09 | 3,88 | 4,28 | 4,31 | 5,17 | 13,79 | 31,9 | 44,83 |
| **Humanities** | 162 | 3,85 | 1,19 | 3,67 | 4,04 | 6,79 | 5,56 | 20,99 | 29,01 | 37,65 |
| **Agricultural science** | 67 | 4,21 | 0,96 | 3,97 | 4,44 | 2,99 | 2,99 | 10,45 | 37,31 | 46,27 |
| *Total* | *1446* | *4,15* | *1,02* | *4,10* | *4,21* | *2,97* | *4,98* | *12,66* | *32,37* | *47,03* |

| Importance: Fast Publishing Process | | | | | | Responses in percent: | | | | Item-Responserate: 92% |
|---|---|---|---|---|---|---|---|---|---|---|
| | Obs. | Mean | Std. Dev. | [95% Conf. Interval] | | Not important | 2 | 3 | 4 | Very important |
| **Natural science** | 483 | 3,49 | 1,05 | 3,39 | 3,58 | 4,14 | 13,04 | 29,61 | 36,23 | 16,98 |
| **Social science** | 442 | 3,41 | 0,97 | 3,32 | 3,50 | 4,75 | 9,73 | 35,97 | 39,14 | 10,41 |
| **Human science** | 169 | 3,52 | 1,00 | 3,37 | 3,67 | 3,55 | 11,24 | 30,18 | 39,64 | 15,38 |
| **Engineering** | 117 | 3,32 | 1,14 | 3,11 | 3,53 | 11,11 | 8,55 | 29,91 | 38,46 | 11,97 |
| **Humanities** | 162 | 3,46 | 1,09 | 3,29 | 3,63 | 4,32 | 14,81 | 30,25 | 31,48 | 19,14 |
| **Agricultural science** | 67 | 3,69 | 0,89 | 3,47 | 3,90 | 1,49 | 4,48 | 37,31 | 37,31 | 19,4 |
| *Total* | *1440* | *3,46* | *1,03* | *3,41* | *3,51* | *4,72* | *11,25* | *32,08* | *37,22* | *14,72* |